\documentclass[preprint]{elsarticle}

\usepackage{lineno,hyperref}
\modulolinenumbers[5]

\journal{Journal of Magnetism and Magnetic Materials}

\begin{document}

\begin{frontmatter}

\title{Thermally controlled confinement of spin wave field in a magnonic YIG waveguide }

\author[ICAT]{Pablo Borys\corref{firstauthor}}
\cortext[firstauthor]{Corresponding author}
\ead{pabloborys@ciencias.unam.mx}

\author[ICAT]{O. Kolokoltsev}
\author[CATEDRAS]{Iv\'an G\'omez-Arista}
\author[KIEV]{V. Zavislyak}
\author[KIEV]{G. A. Melkov}
\author[ICAT]{N. Qureshi}
\author[IF]{César L. Ordóñez-Romero}

\address[ICAT]{Instituto de Ciencias Aplicadas y Tecnología, Universidad Nacional Autónoma de México (UNAM), Ciudad Universitaria, 04510, México}
\address[CATEDRAS]{Cátedras Conacyt – Instituto Nacional de Astrofísica, Óptica y Electrónica, 72840, México}
\address[KIEV]{Faculty of Radiophysics, Electronics and Computer Systems, Taras Shevchenko National University of Kiev, Ukraine}
\address[IF]{Instituto de Física, Universidad Nacional Autónoma de México, Ciudad Universitaria, 04510, México.}

\begin{abstract}
Methods for detecting spin waves rely on electrodynamical coupling between the spin wave dipolar field and an inductive probe. While this coupling is usually treated as constant, in this work, we experimentally and theoretically show that it is indeed temperature dependent. By measuring the spin wave magnetic field as a function of temperature of, and distance to the sample, we demonstrate that there is both a longitudinal and transversal confinement of the field near the YIG-Air interface. Our results are relevant for spin wave detection, in particular in the field of spin wave caloritronics. 
\end{abstract}


\end{frontmatter}

\linenumbers

\section{Introduction}

It is expected that the emergence of thin film logic elements based on spin waves in thin-film ferromagnetic solids can lead to a new generation of Boolean and analogue processors \cite{khitun2008,khitun2010magnonic,kolokoltsev2011synthesis}. One of the important points here is the technique of spin wave excitation and modulation of their parameters. Traditionally, spin waves have been excited and detected using the inductive coupling of micro-electrodes to the dipolar magnetic field of the spin wave system. Usually, this electrodynamical coupling is considered to be constant, however, as shown in this work, it can suffer significant variations, depending on the temperature of the ferromagnetic material. The temperature of the sample can change due to spin wave dissipation, from 1 to 10 $^o$C \cite{kolokoltsev2007optical,fetisov2004thermal} or up to 100-300 $^o$C because of external heating used to control the spin wave propagation \cite{rousseau2015realization,kolokoltsev2012hot}. Recently, the typical electrodynamical and magneto-optical methods for spin wave detection/excitation were enriched with the spin transfer torque (STT) \cite{slonczewski1996current,saitoh2006conversion,kajiwara2010transmission} in Pt/magnet thin film structures caused by electrical or thermal spin currents \cite{wang2011electric,kajiwara2013spin,uchida2008observation,uchida2010observation,uchida2010spin}. STT has been recognized to be a much promising tool to detect exchange SW, to control dipole SW, and to generate thermo-electricity on the basis of spin Seebeck effect. The discovery of the later has stimulated a number of ideas involving magnetocaloritronics \cite{bauer2012spin}. For example, thermally assisted STT has been used for enhancement of spin oscillations in resonators, spin wave amplification and spin auto-oscillations \cite{safranski2017spin,chang2017role,demidov2011wide,lu2012control}. The aim of this work is to reveal lateral effects of sample heating in experimental configurations on the inductive coupling between micro-antennas and the dipolar magnetic field of a spin wave system \cite{an2013unidirectional,jungfleisch2013heat,obry2012spin,an2014control,padron2012amplification,evelt2016high,gladii2016spin,jungfleisch2015thickness,langner2018spin}.
\section{Experiment}

\begin{figure}
	\includegraphics[width=5in]{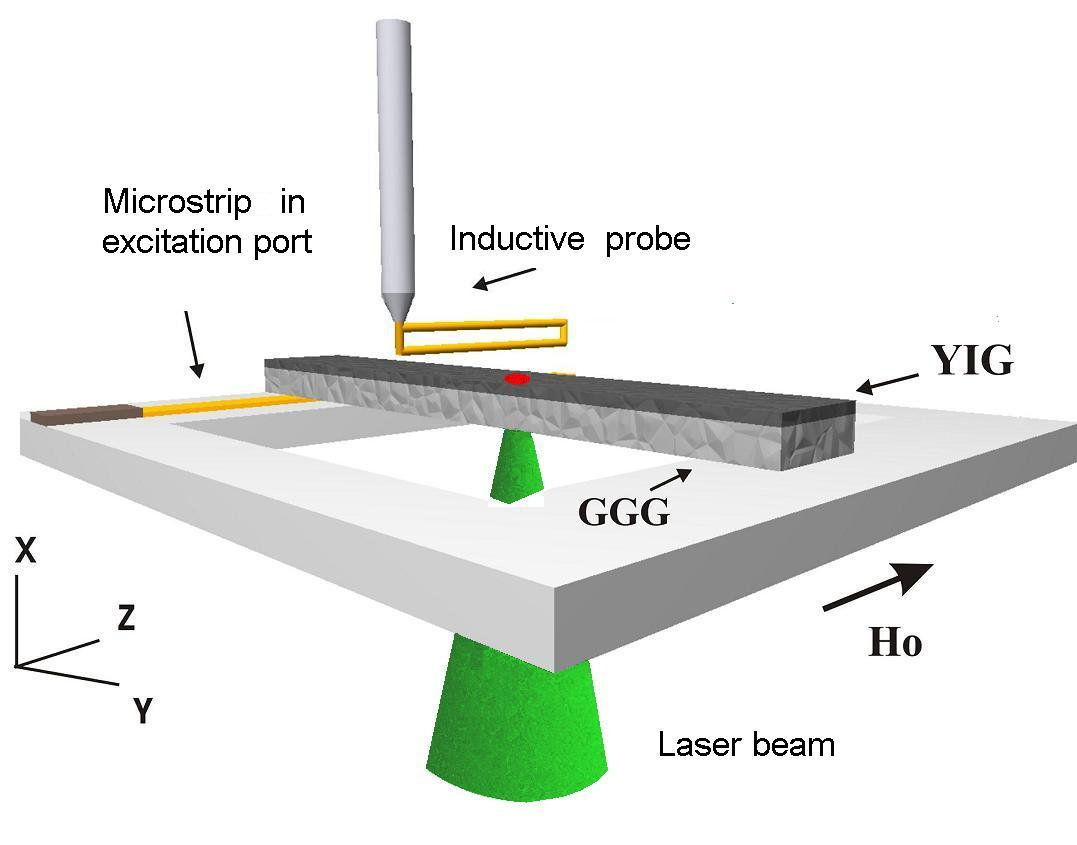}% Here is how to import EPS art
	\caption{A schematic view of experimental set-up. The inductive probe is attached to YX motorized translation stages}	
	\label{fig:1}	
\end{figure}

A schematic diagram of the experimental set-up, designed to investigate the coupling between spin waves propagating on a YIG/GGG sample and  an inductive micro-transducer, is shown in Fig. \ref{fig:1}.  The sample is 1 mm wide in the $Z$ direction and 28 mm long in the $Y$ direction. The thickness of the YIG film is 7 $\mu$m. The sample was biased by a tangential magnetic field ($\mathbf{H}_0$) applied along the $Z$ axis to provide the propagation of Magneto-Static Surface Waves (MSSW) in the $Y$ direction. MSSW were excited at one end of the sample (at $Y=Y_0$), in a pulse regime, by dc electric current pulse flowing through a 0.25 mm-wide microstrip line terminated to a 50 Ohm resistive load. This method provides very short spin wave packets, with duration of $\approx$ 10 ns. In the time domain the shortest period of the magnetization precession in the wave packet is limited by the rise time of the electric current pulse, and in the \textit{k-space} the largest wavenumber (k) is limited by the microstrip line width \cite{kolokoltsev2011synthesis}. The MSSW pulse propagation characteristics were registered by an inductive frame-shaped probe \cite{vlannes1987optical} (Fig. \ref{fig:1}) sensitive to the $Y$ magnetic component of microwave field ($h_y$) induced by the spin wave in the vicinity of the YIG film. The probe was scanned over the sample plane along the $Y$ coordinate (Fig. \ref{fig:1}) by a motorized translation stage. The distance between the probe and the sample surface was also controlled by a motorized translation stage. It should be noted that we used a frame probe with reduced X-dimension to have high spatial resolution of hy along the film normal, as the probe is displaced in the X direction. The probe electrode was fabricated with a 50 $\mu$m micro-wire. The sample was heated with a solid state green laser with variable output power ($P_{opt}$), from 40 to 300 mW. The laser spot on YIG was of 0.5 mm in diameter and was located at the distance of $Y_L$ from the excitation port. 

\begin{figure}
	\includegraphics[width=5in]{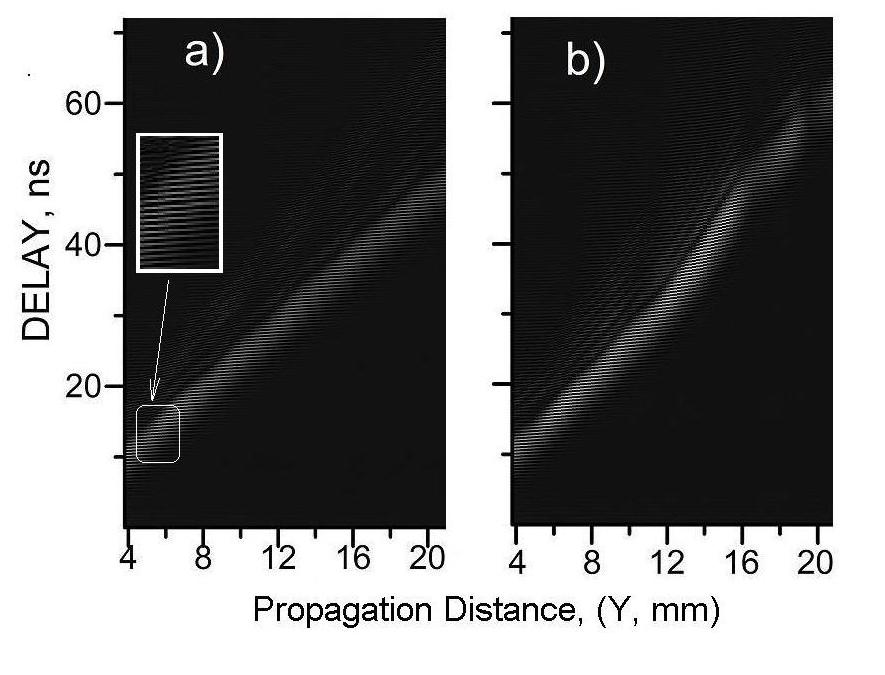}% Here is how to import EPS art
	\caption{Propagation of MSSW pulse along the magnonic waveguide YIG/GGG. The inset in Fig. 2a) shows details the pulse waveform, with a duration of 8 ns, at three adjacent positions along the Y axis. The data represent the pulse waveform (amplitude) a) in the sample at room temperature, and b) in the sample heated at 380 K in its center.}	
	\label{fig:2}	
\end{figure}

Fig \ref{fig:2} compares the time-space evolution of the amplitude of the hy pulse at room temperature (RT sample) in 2(a), with a sample heated at optical power $P_{opt}$ = 180 mW in 2(b). The pulse waveform was recorded by a real time Tektronix oscilloscope with 6 GHz- bandwidth, at different Y- positions of the probe, and at fixed distance $\Delta X=50\mu$m between the probe and the YIG film plane. The measurements were done with a uniform bias field $H_0$ = 120 Oe, and the laser spot at the position $Y_L$= 15 mm. As seen in Fig.\ref{fig:2}, the wave packet in the optically heated sample acquires an additional group delay, compared to the sample at room temperature. This phenomenon has been discussed in ref. \cite{algra1982temperature}, and is caused by a reduction on the saturation magnetization $M_s$ that in turn decreases the slope in the MSSW dispersion relation. 

\begin{figure}
	\includegraphics[width=5in]{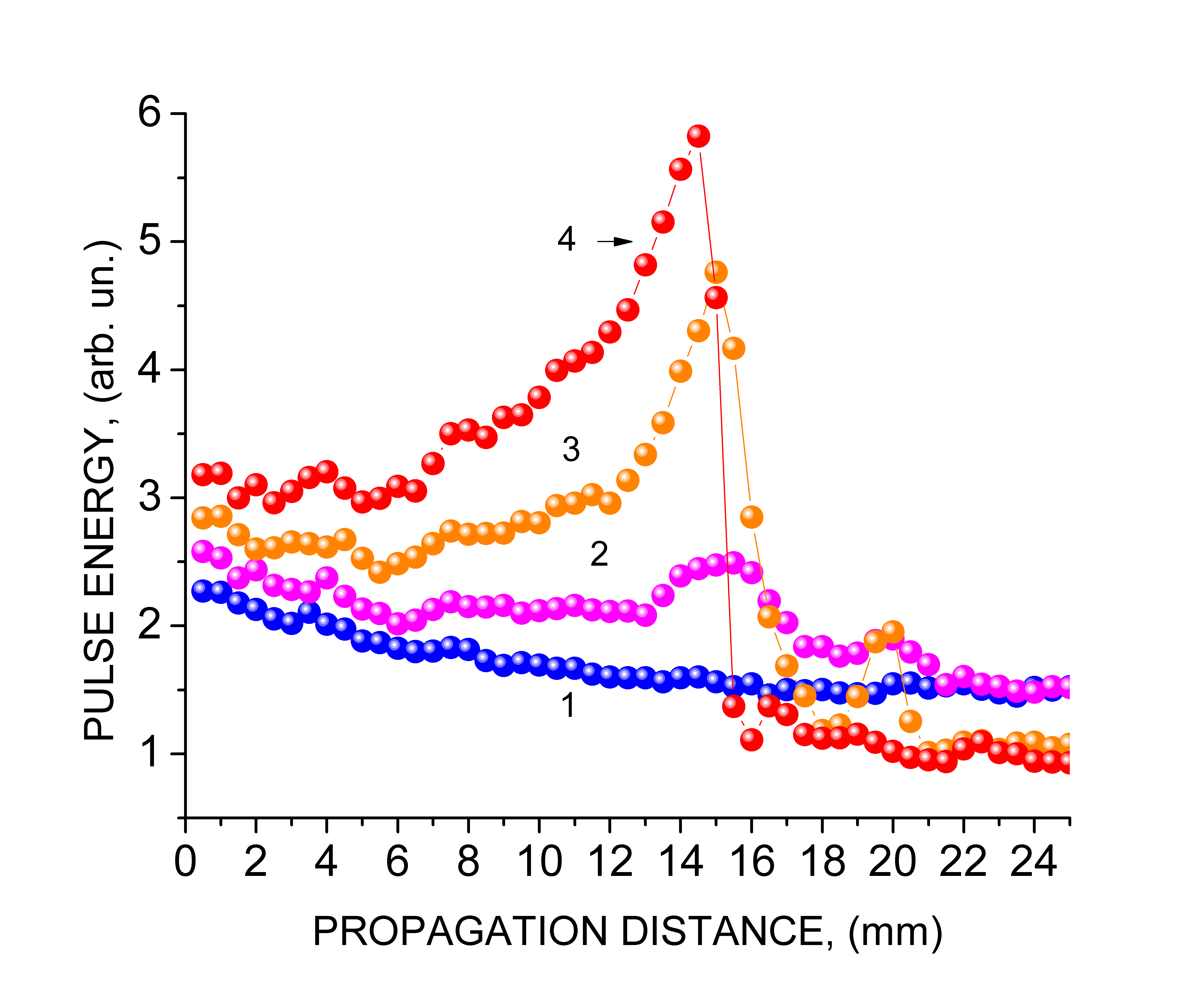}% Here is how to import EPS art
	\caption{Fig. 3 The energy of the MSSW pulse at different distances from excitation port. The curves were recorded at different Popt, which induce different temperatures in the region $Y=Y_L$: 1) T = T$_{ROOM}$; 2) T = T$_{ROOM}$ + 50 K ; 3) T = T$_{ROOM}$ + 70 K ; 4) T = T$_{ROOM}$ + 90 K.}	
	\label{fig:3}	
\end{figure}

\begin{figure}
	\includegraphics[width=5in]{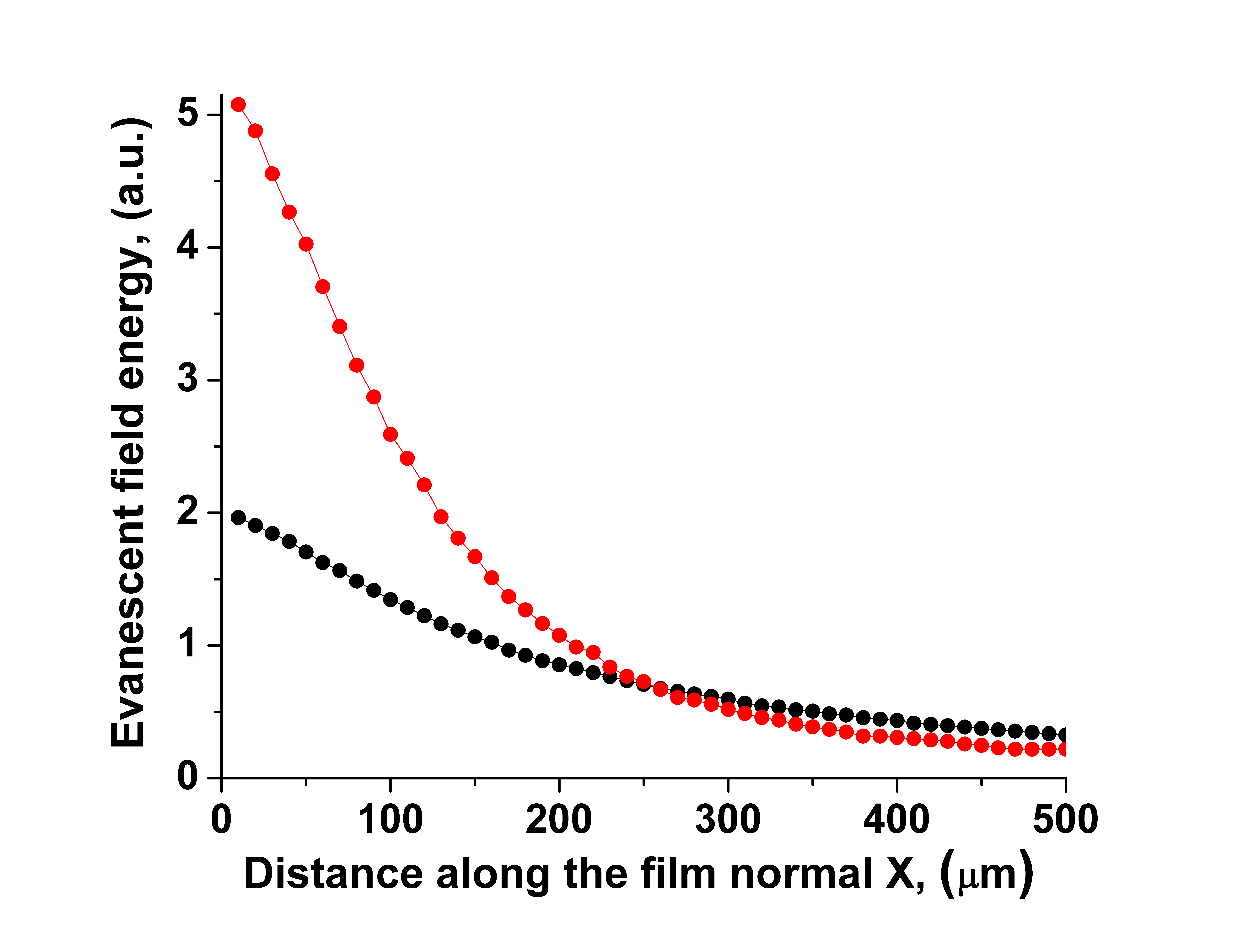}% Here is how to import EPS art
	\caption{Energy of $h_y$ component as a function of the distance between the probe and YIG film surface, at the fixed Y-position of the probe $Y=Y_L$. Red and black experimental points show the energy density in evanescent MSSW field in the heated sample (at T = 380 K) and the RT sample, respectively.}	
	\label{fig:4}	
\end{figure}

Fig. \ref{fig:3} shows the signal detected by the probe, as the probe moves along the Y axis, at different $P_{opt}$. The value of each point in the curves in Fig. \ref{fig:3} represents the energy of the pulse envelope. As clearly seen in the figure, the signal induced in the probe increases in the vicinity of the laser spot, and this increment is proportional to temperature of the hot zone. On the other hand, in the sample at room temperature (Curve 1) the MSSW pulse propagates and attenuates exponentially, in the usual way. The data presented in Fig.3 are proportional to the overlap integral between a small effective area of the probe frame and an evanescent function $h_y(x)$ \cite{vlannes1987optical}. Hence, displacing the probe along the film normal one can obtain the profile $|h_y(x)|^2$, shown in Fig.\ref{fig:4}. In this experiment the probe was located in the center of the hot zone $Y=Y_L$

Fig.\ref{fig:4} presents the principal result of this study: the density of $h_y$ near the film interface increases as the sample temperature increases, i.e. the heat modifies the field confinement.

\section{Theoretical Background}

\begin{figure}
	\includegraphics[width=5in]{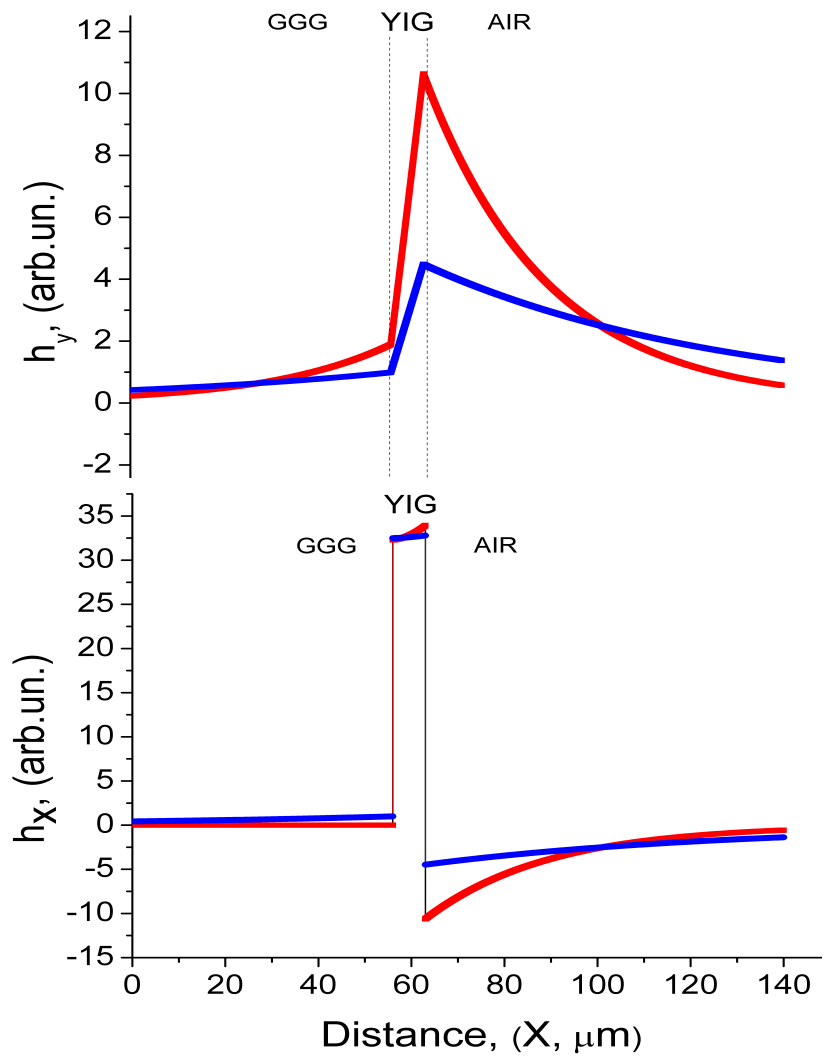}% Here is how to import EPS art
	\caption{Fig.5. The values of the tangential (hy) an the normal (hx) field components calculated for hot (red curves) and RT (blue curves) samples. }	
	\label{fig:5}	
\end{figure}

The effect of the thermally dependent field confinement is caused by the decrease of $M_s$ in the ferrite film, as its temperature increases. It can be analyzed analytically by a full set of Maxwell equations. In our case, considering that the sample is infinite in YZ plane, the solutions for the magnetic and electric fields of MSSW are $\mathbf{h} =(h_x, h_y, 0)$ and $\mathbf{e} = (0, 0, e_z)$, respectively. Let us compare transversal profiles of monochromatic magnetic field components $h_x,\;h_y$ in hot and RT samples, taking into account that the fields have to be normalized to transmit a given power flow $\mathbf{P}$ through the sample. It is clear that in both “hot” and RT samples a value of $\mathbf{P}$ should be the same, supposing equal excitation efficiency of MSSW. It can be shown that the Pointing vector for MSSW is calculated as:

\begin{equation}\label{eq:1}
\mathbf{P}=\frac{c}{8 \pi}\left[-e_{z} h_{y}^{*} \mathbf{i}+e_{z} h_{x}^{*} \mathbf{j}\right]
\end{equation}
or
\begin{equation}
\mathbf{P}_{2}=\frac{c}{8 \pi k_{0} \mu_{\perp}}\left[k e_{z}^{2}+\frac{\mu_{a}}{\mu} e_{z} \frac{\partial e_{z}^{*}}{\partial x}\right] \mathbf{j}
\end{equation}
in the YIG film, and
\begin{equation}
\mathbf{P}_{1,3}=\frac{c}{8 \pi k_{0}} k e_{z}^{2} \mathbf{j}
\end{equation}
in air and substrate.\\

Here: $k$ is the MSSW wavenumber, $k_0=\omega/c$, $c$ is the speed of light in the vacuum, $\omega$ is the MSSW frequency, $\mu=(\omega^2-\omega_1^2)/(\omega^2-\omega_H^2)$, $\mu_a=\omega\omega_M/(\omega^2-\omega_H^2)$, $\omega_H=\gamma H_0$, and $\omega_M=4\pi M_S$, $\omega_1=\omega_H(\omega+\omega_M)$, and $\gamma$ is the electron gyromagnetic ratio.

Then, taking into account that $\mathbf{h}$, $\mathbf{e}$ in Eq. \ref{eq:1} are proportional to a certain constant, $A$, the value of $A$ for both hot and RT sample can be calculated using the condition $\sum_{i=1,2,3}\mathbf{P}_i(Hot\;sample)=\sum_{i=1,2,3}\mathbf{P}_i(RT\;sample)=Const$. The explicit expressions for MSSW field components in Eq. \ref{eq:1} are given in Appendix A. The calculated magnetic field profiles are shown in Fig. \ref{fig:5}.

The results were obtained by using the experimental approximation for temperature dependence of the saturation magnetization in YIG:$M_s=140-\alpha\Delta T (G)$, $\alpha\approx0.3$ G/K \cite{algra1982temperature}. The field profiles in Fig.\ref{fig:5} correlate well with the experimental profiles in Fig.\ref{fig:4}. 

\section{Discussion and Conclusions}

The peculiarity of the results for the pulse group delay shown in Fig. \ref{fig:2} is that the local heating increases the pulse delay, however, it does not change the group velocity dispersion. As seen in Fig.\ref{fig:2}, the pulse width (the pulse duration) in the hot region remains unchanged, with respect to the pulse width in the RT sample. This means that spatial width of the pulse along the Y coordinate decreases, i.e. there is spatial, longitudinal compression of the pulse along the propagation direction. This leads to the increase of a peak and average amplitude of the pulse envelope for pulse power to be conserved. The effect has been analyzed in \cite{kolokoltsev2015compression}, where we used a large diameter loop antenna that was not sensitive to the effect of the transversal confinement of the evanescent field shown in Fig.\ref{fig:4}, and \ref{fig:5}.
On the other hand, the results presented in Fig.\ref{fig:3}, \ref{fig:4}, and \ref{fig:5} indicate that increasing the sample temperature increases the coupling between MSSW field and the micro-antenna. The experimental, Fig.\ref{fig:4}, and theoretical, Fig.\ref{fig:5}, data demonstrate that this effect takes place due to an increasing concentration of magnetic fields near YIG-Air interface, the so-called transversal confinement. 
In conclusion, it is shown that the increase of the sample temperature leads to the increase of both longitudinal and transversal confinement of MSSW in the vecinity of YIG film. This effect, in turn, is revealed as the increase of the signal induced in a micro-antenna,  that has to be taken into account in the experiments on spin-wave caloritronics.

\section{Appendix A}

Full system of Maxwell equations for electromagnetic waves in the sample saturated in the Z direction describes two kind of waves. The subsystem

\begin{equation}
\begin{array}{c}{\frac{\partial h_{z}}{\partial y}=i \varepsilon k_{0} e_{x}} \\ {\frac{\partial h_{z}}{\partial y}=-i \varepsilon k_{0} e_{y}} \\ {\frac{\partial e_{y}}{\partial x}-\frac{\partial e_{x}}{\partial y}=-i k_{0} h_{z}}\end{array}
\end{equation}

describes fast waves, which neglects magnetism, and the subsystem 

\begin{equation}
\begin{array}{c}{\frac{\partial h_{y}}{\partial x}-\frac{\partial h_{x}}{\partial y}=i \varepsilon k_{0} e_{z}} \\ {-i k_{0}\left(\mu h_{x}-i \mu_{a} h_{y}\right)=\frac{\partial e_{z}}{\partial y}} \\ {i k_{0}\left(i \mu_{a} h_{x}+\mu h_{y}\right)=\frac{\partial e_{z}}{\partial y}}\end{array}
\end{equation}

that is used to describe MSSW. It can be reduced to

\begin{equation}
\frac{\partial^{2} e_{z}}{\partial x^{2}}+\frac{\partial^{2} e_{z}}{\partial y^{2}}+\varepsilon \mu_{\perp} k_{0}^{2} e_{z}=0
\end{equation}

MSSW fields, where $\mu_{\perp}=(\mu^2-\mu_a^2)/\mu$, which satisfy Eq. 2b, and Eq.3 are

\begin{equation}
\underbrace{e_{z}=A e^{\beta_{a} x+i(\omega t-k y)}, h_{x}=\frac{k}{k_{0}} A e^{\beta_{a} x+i(\omega t-k y)}, h_{y}=-i \frac{\beta_{a}}{k_{0}} A e^{\beta_{a} x+i(\omega t-k y)}}_\textbf{Air}
\end{equation}

\begin{equation}
\underbrace{\begin{array}{l}{e_{z}=\left(B \cosh\left(\beta_{m} x\right)+C\sinh\left(\beta_{m} x\right)\right) e^{i(\omega t-k y)}} \\ {h_{x}=\frac{1}{k_{0}\left(\mu^{2}-\mu_{a}^{2}\right)}\left[\mu k\left(B \cosh\left(\beta_{m} x\right)+C\sinh\left(\beta_{m} x\right)\right)+\mu_{a} \beta_{m}\left(B \sinh\left(\beta_{m} x\right)+C\cosh\left(\beta_{m} x\right)\right)\right] e^{i(\omega t-k y)}} \\ {h_{y}=\frac{1}{k_{0}\left(\mu^{2}-\mu_{a}^{2}\right)}\left[\mu_{a} k\left(B \cosh\left(\beta_{m} x\right)+C\sinh\left(\beta_{m} x\right)\right)+\mu \beta_{m}\left(B \sinh\left(\beta_{m} x\right)+C\cosh\left(\beta_{m} x\right)\right)\right] e^{i(\omega t-k y)}}\end{array}}_\textbf{YIG}
\end{equation}

\begin{equation}
\underbrace{
e_{z}=D e^{-\beta_{a} x+i(\omega t-k y)}, \quad h_{x}=\frac{k}{k_{0}} D e^{-\beta_{a} x+i(\omega t-k y)}, \quad h_{y}=i \frac{\beta_{a}}{k_{0}} D e^{-\beta_{a} x+i(\omega t-k y)},}_\textbf{Substrate}
\end{equation}

with $\beta_{a}=\sqrt{k^2-k_0^2}$, and $\beta_{m}=\sqrt{k^2-\epsilon\mu_{\perp}k_0^2}$. The standard electrodynamic boundary conditions at the structure interfaces determine the following relations between the coefficients A,B,C,D

\begin{equation}
\begin{array}{l}{A=B, \quad D e^{-\beta_{a} s}=B \cosh\left(\beta_{m} s\right)+C\cosh\left(\beta_{m} s\right), \quad \beta_{a}\left(\mu^{2}-\mu_{a}^{2}\right) A=\mu_{a} k B+\mu \beta_{m} C}, \\ {-\beta_{a}\left(\mu^{2}-\mu_{a}^{2}\right) e^{-\beta_{a} s} D=\mu_{a} k\left(B \cosh\left(\beta_{m} s\right)+C\sinh\left(\beta_{m} s\right)\right)+\mu \beta_{m}\left(B\sinh\left(\beta_{m} s\right)+C\cosh\left(\beta_{m} s\right)\right)}\end{array}
\end{equation}

Then, the constant A is calculated from the condition

\begin{equation}
\sum_{i=1,2,3}\mathbf{P}_i(Hot\;sample)=\sum_{i=1,2,3}\mathbf{P}_i(RT\;sample)=Const.
\end{equation}

\section{Acknowledgements}

This work was supported by the UNAM-DGAPA research grant IG100517, and by fellowship BECA UNAM Posdoctoral. Dr. O. Kolokoltsev is thankful to UNAM-DGAPA for sabbatical scholarship.

\bibliography{mybibfile}

\end{document}